\documentclass[showpacs,twocolumn,prl]{revtex4}
\usepackage{amsmath,graphicx}
\usepackage{color,ulem}

\definecolor{darkred}{rgb}{0.90,0,0}
\definecolor{darkgreen}{rgb}{0,0.60,.2}
\definecolor{darkblue}{rgb}{0,0,1}
\definecolor{grey}{cmyk}{0,0,0,0.25}
\definecolor{orange}{cmyk}{0,0.6,0.8,0}

\begin{document}

%void void void void void 

\title {Conduction of DNA molecules attached to a disconnected array of metallic Ga nanoparticles.}

\author{ A.D. Chepelianskii$^1$, D. Klinov$^2$, A. Kasumov$^1$, S. Gu\'eron$^1$, O. Pietrement$^3$, S. Lyonnais$^4$ and H. Bouchiat$^1$}
\affiliation {$^1$ Univ. Paris-Sud, CNRS, UMR 8502, F-91405, Orsay, France \\
$^2$ Shemyakin-Ovchinnikov Institute of Bioorganic Chemistry, Russian Academy of Sciences, Miklukho-Maklaya 16/10, Moscow 117871, Russia \\
$^3$ UMR 8126 CNRS-IGR-UPS, Institut Gustave-Roussy, 39 rue Camille Desmoulins, 94805 Villejuif Cedex, France \\
$^4$ Museum National d'Histoire Naturelle, CNRS, UMR7196, Inserm, U565, 43 rue Cuvier, 75005 Paris, France.}

\begin{abstract}
We have investigated the conduction over a wide range of temperature of $\lambda$ DNA molecules deposited across %insulating 
slits etched  through a few nanometers thick platinum film. The slits are insulating before DNA deposition but contain metallic  Ga nanoparticles, a result of focused ion beam  etching. When these nanoparticles are superconducting we find that they can induce   superconductivity through the DNA molecules, even though the main electrodes are non superconducting. These results indicate 
that minute metallic particles can easily transfer charge carriers  to  attached DNA molecules and provide a possible reconciliation between apparently contradictory previous experimental results concerning the length over which DNA molecules can conduct electricity.
\end{abstract}

\maketitle

%\section{Introduction} 

Conductivity of DNA is a long-standing debate. Following the initial predictions that DNA molecules 
should conduct electricity, several types of experiments were attempted to probe the conduction 
mechanisms, ranging from emission and absorption spectroscopies \cite{barton1} to microwave absorption \cite{rf}. 
Several groups  have also attempted  direct measurements of DNA conductivity 
by attaching DNA molecules to metallic electrodes \cite{porath1,fink}. The contradictory experimental results,
with behaviors ranging from insulator to coherent quantum transport over distances 
in the hundred nanometer range led to a strong controversy \cite{endres}. The picture emerging in the past few 
years has been that DNA can conduct over distances of  tens of nanometers:
this was shown by STM and local probe techniques \cite{porath,elke, bingqian,wang}, as well as in a spectacular experiment \cite{barton} : 
a $3$ nanometer  long DNA molecule was inserted in a cut carbon nanotube,
increasing its initial resistance only twofolds, and was subjected 
to biological manipulations that altered and then restored the conductivity. The importance of the environment of the molecules  in order to have reproducible results was pointed out in \cite{mahapatro09}.
Conduction over hundreds of nanometers, and up to several microns, was  also reported by different groups \cite{okahata,fink, hartzell, heim}, including ours.
In our previous experiments DNA was found to be conductive between platinum-carbon electrodes \cite{klinov} and 
between rhenium-carbon electrodes \cite{kasumov}. In this last case as the samples were lowered 
below the superconducting critical temperature of the electrodes (rhenium is a superconductor with $T_c=1.7~K$) the sample resistance 
decreased, indicating coherent quantum transport through the DNA molecules.
These results are both technologically and fundamentally important since
long range transport in DNA molecules may lead to the creation of new nanoscale self-assembled 
electronic devices. From the fundamental point of view, DNA is one of the rare 
one-dimensional molecular wires that can be obtained in mono dispersed form 
with known chemical structure and chirality. 
It is thus important to understand the ingredients that lead to conduction over %micron 
long distances.

In this Letter we reconcile previous findings by showing that conduction over 
distances greater than hundreds of nanometers can occur if the DNA molecules are attached 
to a disconnected array of nanoparticles (typically 10 to 20 nm apart)  that locally dopes the molecules, enhancing conduction.  
In addition in our case the nanoparticles are superconducting, which induces 
superconducting correlations in DNA at low temperatures.

All our samples, including the previous ones, are fabricated with unconventional techniques:
without electron beam lithography and with functionalization of the 
sample surface by a pentylamine plasma. Pentylamine was used because 
it is known to promote attachment of DNA to amorphous carbon films (such as those used in transmission electron microscopy, see \cite{dubochet}). 
The samples we describe hereafter are also fabricated 
using focused ion beam etching of a thin platinum carbon film 
deposited on mica, with subsequent pentylamine plasma treatment 
before deposition of DNA. 
Compared to our previous experiments we have gained a better understanding 
of this functionalization technique, establishing that pentylamine 
adheres only on carbonated surfaces and not directly on mica or metals. 
Thus fabrication begins with a mica substrate covered by a e-gun deposited 
platinum carbon film a few nanometers thick (5 nm thick,  square  resistance 1 kOhms). 
Although the carbon concentration is not known exactly, 
we checked that the concentration was high enough to anchor the pentylamine,
since no DNA attached to a platinum surface without co-deposited carbon. 
We deposit thick gold contact pads through a mechanical mask and
divide the centimeter square mica substrate into roughly 
twelve sample regions using a UV laser with a 30 micron diameter laser beam,see Fig.1.
We then proceed to etch away the metal over a thin, $50\;{\rm \mu m}$ long, region 
using a focused ion beam (magnification $\times 3000$ and current $3.5\;{\rm pA}$). 
In order to obtain narrow insulating regions we monitor the resistance of a first slit as we 
etch the platinum film one line-scan at a time. We stop the etching as soon as the resistance diverges, see Fig.1.
The other slits are etched using a slightly larger (15\%) number of scans than was necessary to open the 
first slit. We then check electrically with a probe station that all slits have a resistance above a few G$\Omega$. 
The width of the slits fabricated with this technique ranges between 70 and 150 nm.
The next step is pentylamine deposition, in a DC plasma discharge with pentylamine 
vapor pressure $P = 0.1$ Torr,  and current $I= 3-5$ mA for a few minutes.
%voltage $V = $, flow of pentylamine vapor $...$ and 
%plasma duration $ = $. 
A drop of $\lambda$-DNA \cite{dna} solution was incubated on the substrate 
surface for a few minutes and then rinsed away 
using a water flow created by a peristaltic-pump  (flow  few cm/s).
Out of eight mica substrates on which DNA deposition was attempted \cite{orsaymoscou}, 
five were covered by DNA molecules as established by atomic force microscopy. 
These five substrates contained around 30 slits. All samples on two of these substrates 
were completely insulating. On the other three substrates 11 out of 15 samples 
were conducting. We have also   prepared a control substrate, incubated with the same buffer but without DNA molecules, and rinsed like the other samples. We found that all 14 slits  etched on  these samples remained insulating.  
\begin{figure}
\centering
\includegraphics[width= 0.8 \columnwidth]{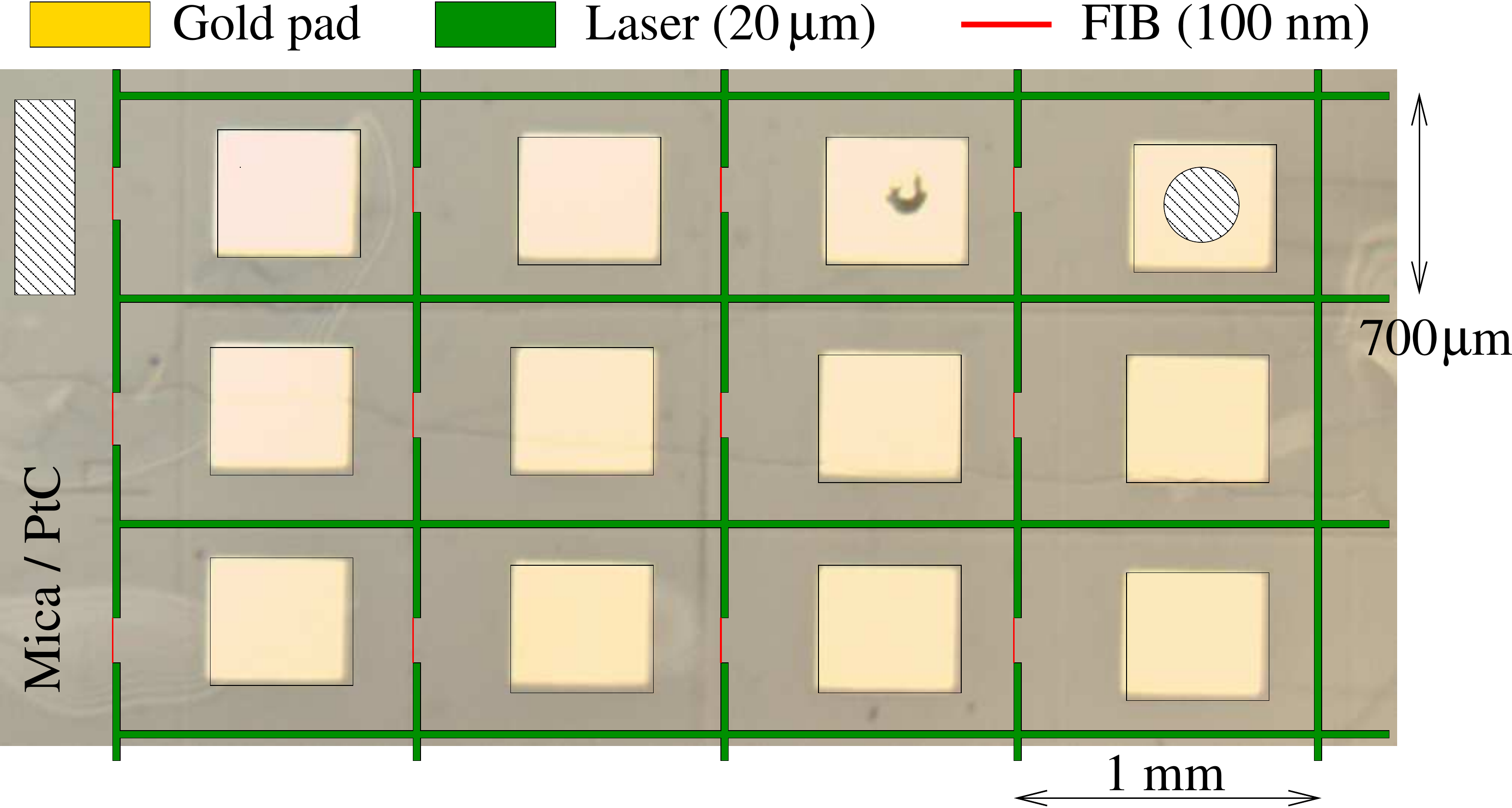}
\includegraphics[width= 0.8 \columnwidth]{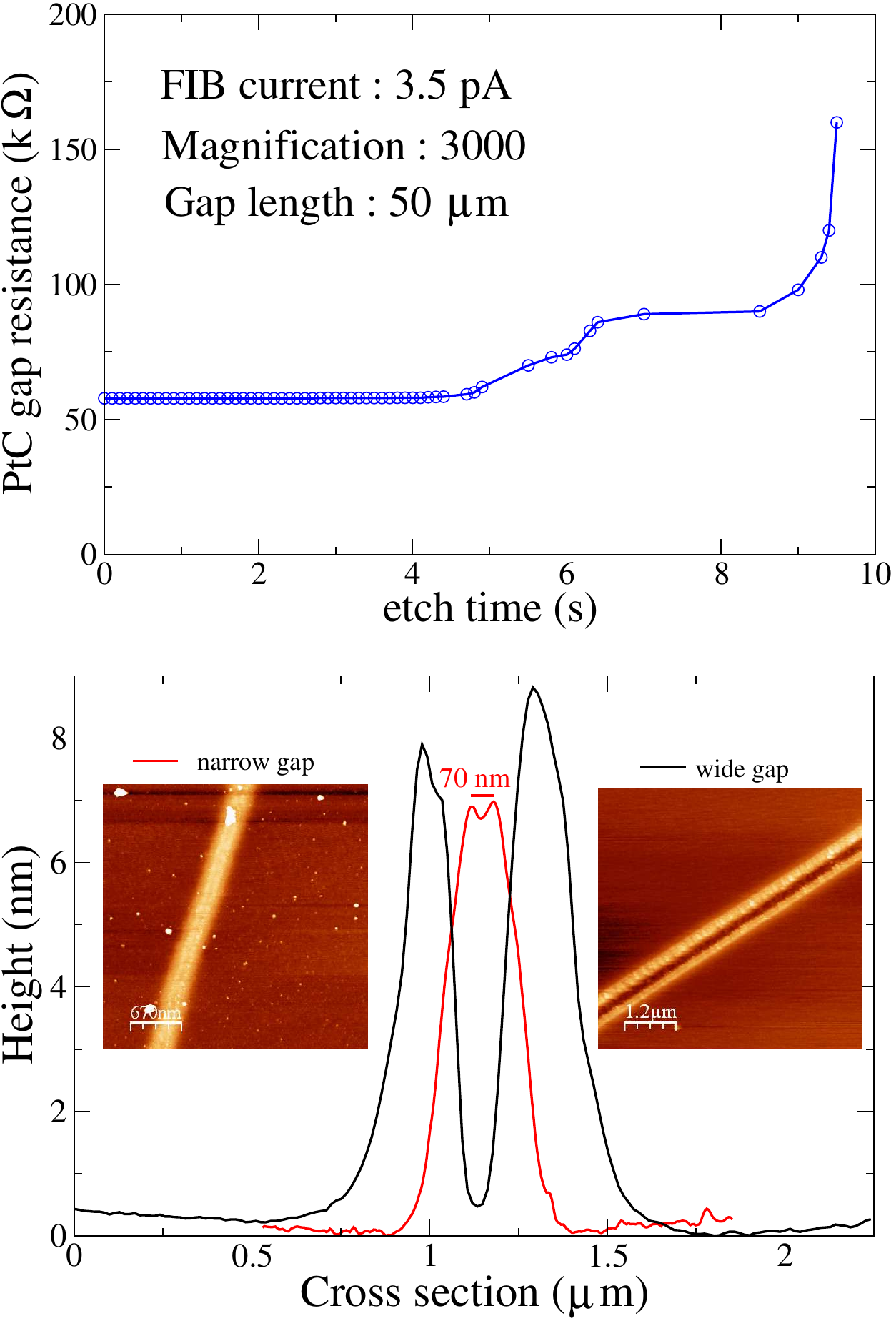}

\caption{Top panel:general view of the samples showing the macroscopic gold pads on the Pt/C film with the laser and FIB etched slits.
Medium panel: Resistance of a gap during FIB etching as a function of exposure time. The gap is etched in a single scan mode 
with a scan time of $0.1\;{\rm s}$ which allows to measure the resistance after each scan. After total time $t > 9.7\;{\rm s}$
the resistance jumps and the gap becomes insulating. Bottom panel: Averaged height profile from two gaps prepared using FIB, their AFM images are shown in the color insets. 
The narrow gap was obtained during the calibration resistance measurement from 
Fig.~\ref{fig:FibEtchTime} while the wide gap was obtained with a larger exposure time. 
The width of the insulating region is hard to measure precisely with AFM because of the residues produced during etching, which accumulate at the edges of the slit. 
}
\label{fig:FibEtchTime}
\end{figure}
Room temperature conductance was measured in a probe station, using an ac voltage in the mV range at frequencies ranging from 1 to 30Hz. The resistance of conducting samples was found to vary, depending on the slit, between 5 k$\Omega$ and 50 k$\Omega$. These values are consistent with previous findings \cite{kasumov,klinov}, given that the number of deposited  molecules across each slit  varies between 10 and 100.  

The pentylamine plasma creates a positively charged organic layer that allows DNA molecules to bind to the carbonated hydrophobic electrodes.  Conducting AFM characterization of this pentylamine layer on a smooth Pt/C film indicates that  the pentylamine film forms a smooth  insulating layer. This  is not the case along  the edge of the  slits, where  FIB etching as well  as  unavoidable carbon contamination  introduce roughness, leading  to defects and holes in the pentylamine coverage. As a result the edges of the slit remain metallic, as is needed to establish electrical contact to the DNA on both sides of the slit.

We have used both atomic force atomic microscopy and  high resolution scanning electron microscopy  to characterize the structure of the FIB etched slit. We find that the insides of the slits are rather rough for two reasons:
The incomplete etching of the platinum film leaves metallic disconnected islands of typical size 10 nanometers. In addition, some slits contain a disordered  array  of  roughly spherical nanoparticles  (see Fig.2). The regular shape of these spheres contrasts  with the  irregular shape of the etching residues of PtC.  As confirmed by transport experiments presented below,  these spherical nanoparticles result from  condensed gallium drops generated by the FIB. Their size varies between 3 and 10 nm, their separation between 5 and 20 nm.   Even if these nanoparticles do not directly contribute to electronic transport through the slits, which were insulating before deposition of DNA and remained so after a flow of saline buffer solution without molecules, we will see that they certainly modify the electronic properties of DNA  molecules deposited  across the slit. 

\begin{figure}[ht]
\centering
\includegraphics[width= \columnwidth]{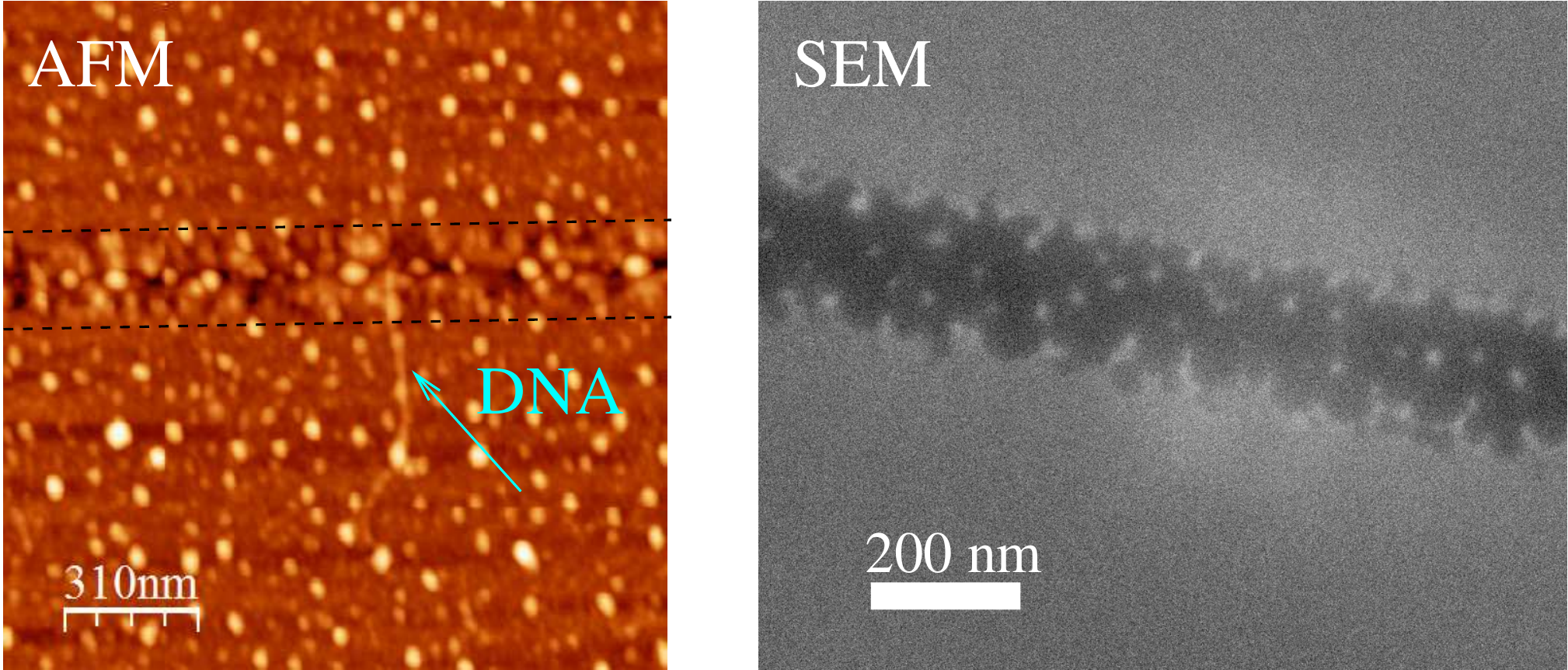}

\caption{
Left panel- Atomic force microscopy  image   of one of the sample where low temperature transport was investigated, taken  using an ultra sharp AFM tip and showing the presence of  a DNA molecule across the slit.
The slit is nearly invisible due to the scanning direction chosen to be parallel to the slit in order to optimize the DNA visualization.
 Right Panel- Electron microscopy image  of the same sample  Gallium nanoparticles are clearly visible in the etched slit region.
%However on typical images DNA molecules seem absent  (see panel c of this figure, or Fig.~\ref{fig:Box2B} which was obtained in Orsay).
\label{fig:Box2Bdima}
}
\end{figure}

In the following we present low temperature transport measurements of DNA molecules deposited through slits decorated with gallium nanoparticles.
The samples investigated have resistances ranging from 5 to 20 $ k\Omega$ at room temperature, with roughly 10 to 30 connected molecules, as deduced from the density of molecules on the substrate far from the slit. The samples were electronically and mechanically connected by gold plated spring contacts \cite{fragile} on the gold pads on the Pt/C film, and mounted in a dilution refrigerator operating down to 50 mK.  The resistance was measured  via lines with room temperature low pass filters. Measurements were performed in a current biased configuration using an ac current source of 1 nA operating at 27 Hz and a Lock-in detector with a low noise voltage pre-amplifier. Whereas the resistance was nearly independent of temperature between room temperature and 4 K, it dropped as T decreased, with a broad transition to a value of the order of 4 $ k\Omega$ (which corresponds to the resistance of the normal Pt/C electrodes in series with the DNA molecules), see Fig. 3.  This transition to partial proximity-induced superconductivity is shifted to lower temperatures in a magnetic field.  It is the broadest for the most resistive sample, and exhibits the smallest magnetic field dependence. 

\begin{figure}
\centering
\includegraphics[width= 0.8 \columnwidth]{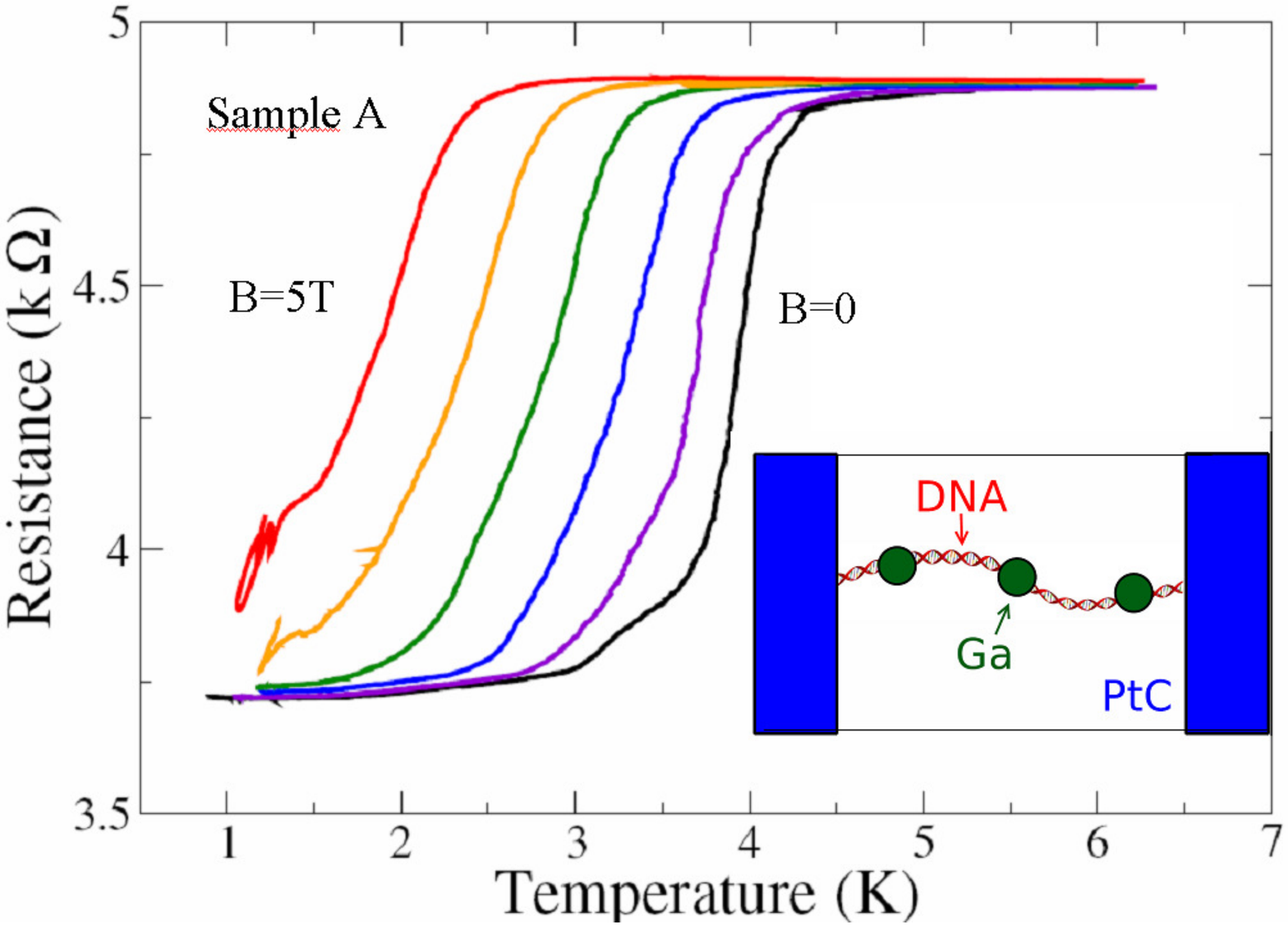}
\includegraphics[width= 0.8 \columnwidth]{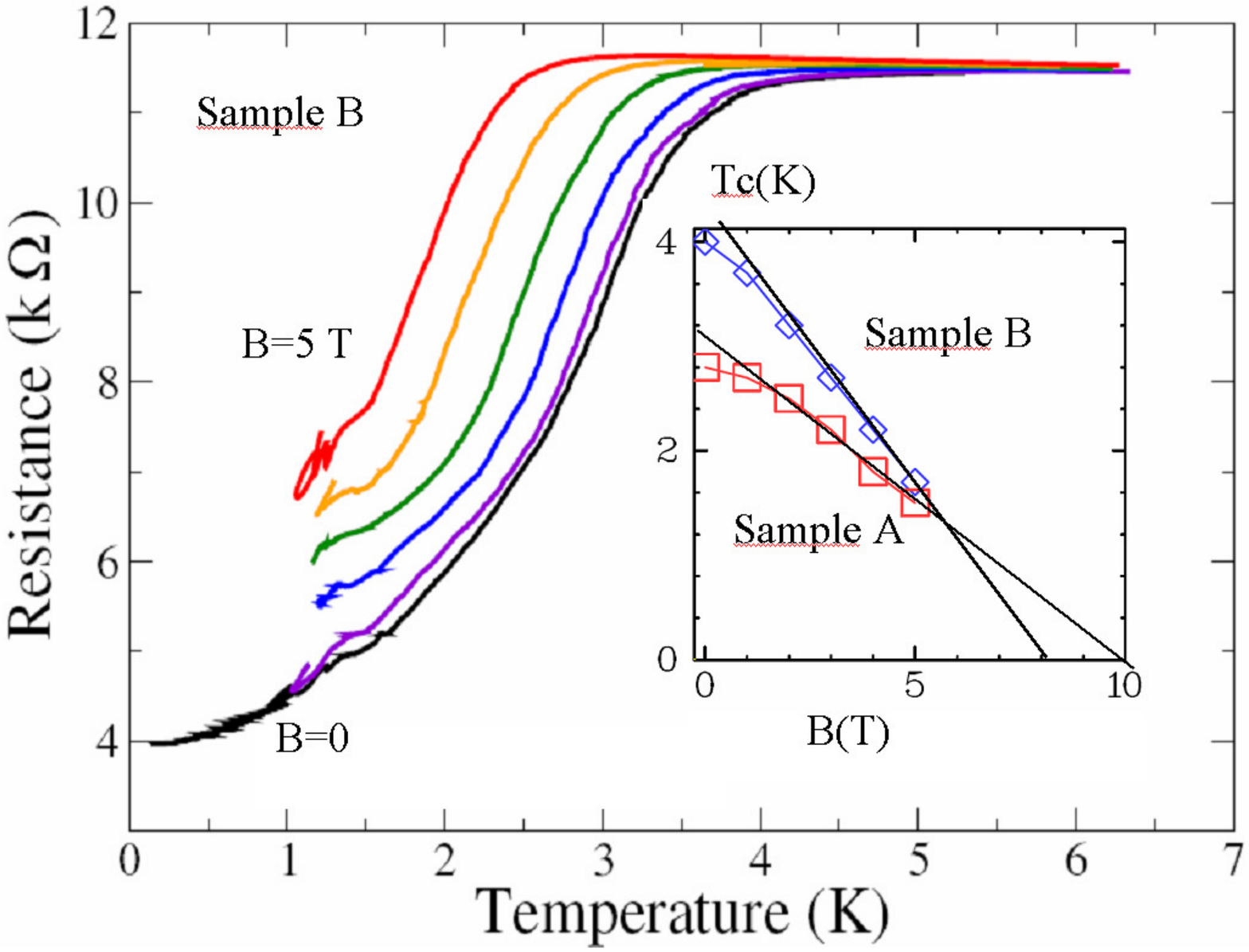}

\caption{%Top panel: Scheme of the mechanical connection system we used to contact our DNA samples. 
 Low temperature dependence at several magnetic fields (going from O,1  to 5T ) of the resistance for 2 different samples where Ga nanoparticles are present inside the slit as described in the inset of the top panel.
 Inset of bottom panel: magnetic field dependence of the critical temperature $T_c(H)$ deduced from the inflexion points of  the $R(T)$ curves.
%Bottom panel: shows the temperature dependence for the $4.8\;{\rm k \Omega}$ junction.
}
\label{fig:DnaNanoFig2}
\end{figure}

Another superconducting-like feature is the non linear IV curves at low temperature, see Fig. 4: The  dc current-dependent differential resistance is lowest at small dc current and increases with increasing dc current. The increase is non monotonous,  presenting several peaks up to a current of the order of $1\mu A$, a sort of critical current, above which the resistance is constant and independent of dc current.
 The many peaks in the differential resistance curves are typical of non homogeneous superconductivity. For instance the differential resistance jumps seen in narrow superconducting wires (diameter smaller than coherence length) are associated with the weak spots of the wire. Since  neither the Pt/C electrodes  nor the DNA molecules are superconducting (as shown in previous experiments),  these results suggest that the gallium nanoparticles, which are superconducting, induce superconductivity through the DNA molecules.  
The superconducting transition temperature of pure gallium is $T_c=1~K$ but it is reasonable to expect that the  gallium nanoparticles,  because of  their small size and  their probable large carbon content, have a higher $T_c$ \cite{tcgallium}.   It is interesting to note that the low intrinsic carrier density in the DNA molecules may prevent the  inverse proximity effect, i.e. the destruction of the superconductivity  of the gallium nanoparticles.  Those same nanoparticles could not induce any proximity effect in metallic wires because of the high density of carriers in metals. This possibility of inducing long range superconductivity with superconducting nanoparticles was investigated  recently in the context of graphene \cite{kostya}.   In the present case, it is also possible that the gallium nanoparticles could contribute to carrier doping of the DNA molecules in the normal state.
\begin{figure}
\centering
\includegraphics[width=\columnwidth]{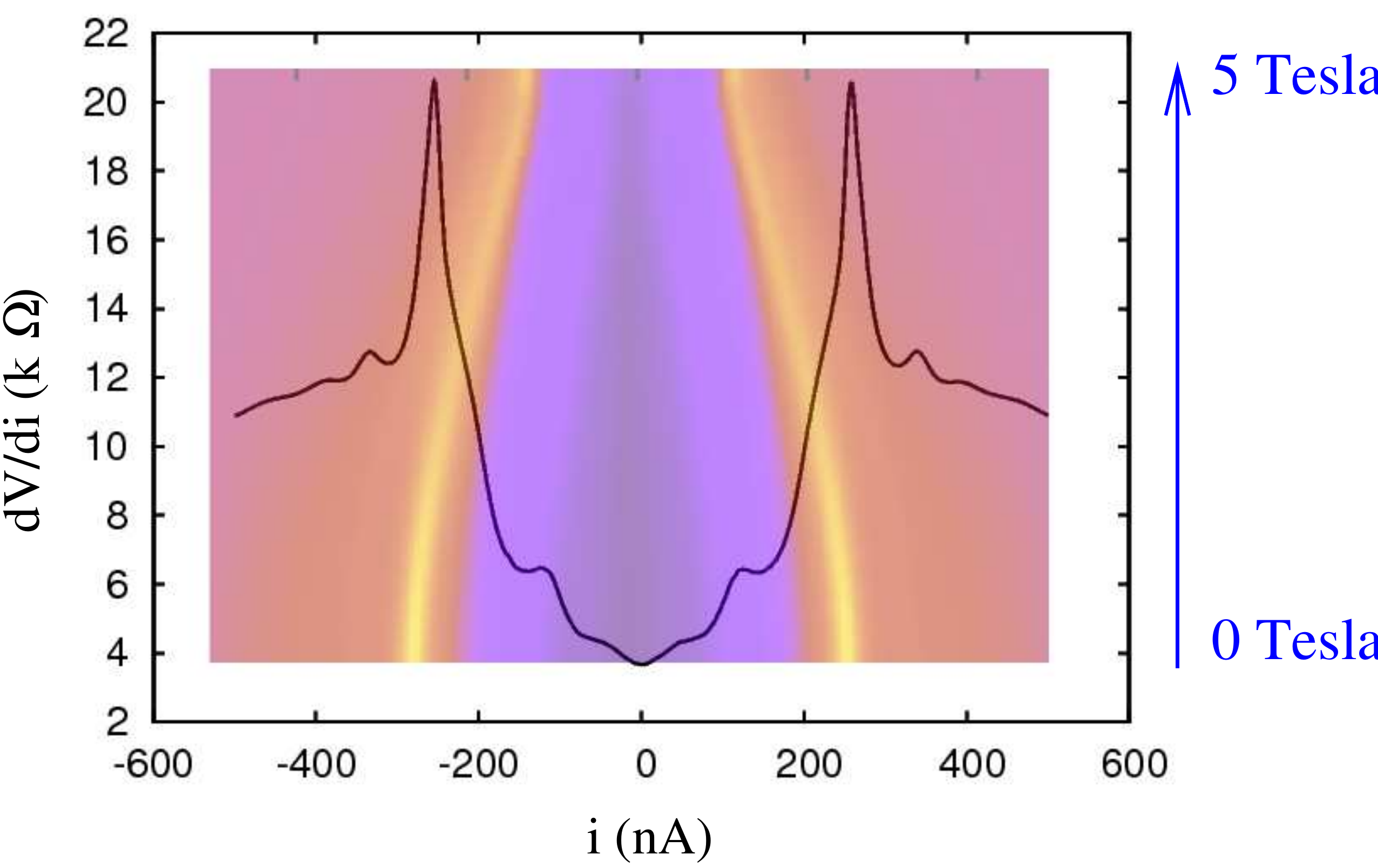}
\caption{The black curve represents the differential resistance $dV/di$ as a function of DC current through the $10\;{\rm k\Omega}$ 
sample at $100\;{\rm mK}$. The color inset in the background shows the evolution of the differential resistance
encoded as a color scale with yellow/violet representing maximal/minimal differential resistance. 
The $x$ axis represents the DC-current as in the main figure, and the $y$ axis indicates the magnetic field 
ranging from $0$ to $5$ Tesla. 
}
\label{fig:DnaNanoFig3}
\end{figure}

The difference between the  transitions of the various samples is probably related to the existence of nanoparticles  of different sizes, leading to  superconducting transitions more rounded and with a weaker $T_C(H)$ dependence in small particles than in large ones.  The radius $R$ of the nanoparticles inducing superconductivity  in DNA can be estimated from the critical field $H_c=\Phi_0/\pi R^2$, for which the transition temperature extrapolates to 0. This field (see Fig.3) is of the order of  10 T, corresponding to a radius between 5 and 7 nm. A rough estimate of the number of nanoparticles bound to DNA molecules participating in transport can  also be deduced from the number of peaks of differential resistance which varies from 3 to 6 depending on the samples (the largest number of peaks is observed in the lowest resistance samples). This corresponds to a typical distance between nanoparticles attached to a DNA molecule of the order of 10 to 20 nm, which is thus the length over which we probe electronic transport along the DNA molecules, and not the total width of the slit. The relatively low values of measured resistances, as well as the appearance of proximity induced superconductivity, indicates a strong electronic coupling between the DNA molecules and  both Pt/C residues and Ga nanoparticles. This contrasts with previous measurements of DNA molecules linking gold nanoparticles  \cite{nanoau}, where the conductivity did not exceed $10^{-4}$ Scm$^{-1}$ for a distance between metallic nanoparticles of 10 nm, whereas the conductivity in the present case can be estimated to be  of the order of unity in the same units.  Accordingly transport experiments on completely metallised DNA molecule\cite{sivan} did not seem to indicate any intrinsic contribution of the DNA molecules to the conduction measured.  These differences may originate in the nature of the binding between the metallic nanoparticles and the DNA which in ref. \cite{nanoau} was of covalent nature (involving alkanethiol  molecules of low conductivity), whereas in the present case we believe that a good electrical contact between DNA molecules and the metallic nanoparticles is provided by the discontinuities and defects in the pentylamine film.    

Our  results indicate that minute metallic particles  can transfer charge carriers  to  attached DNA molecules and confirm that DNA molecules can be conducting on lengths of the order of 10 nm but we cannot conclude with these experiments on the conduction on longer length scales.  Since in our previous experiments \cite{kasumov, klinov} the DNA molecules were connected across similarly etched slits in thin metallic films, the existence of metallic residues cannot be excluded, and the conduction of DNA molecules could thus also have been probed on distances no greater than 10 nm. These results invite to a systematic investigation of the possible carrier doping of DNA by metallic nanoparticles.

We thank F. Livolant, A. Leforestier,  D. Vuillaume  and D. Deresmes for fruitful discussions and acknowledge ANR QuantADN and DGA for support.

%Sample fabrication results from our progress in understanding the role of the 
%pentylamine functionalization of the sample surface. 

%\section{Sample Fabrication} 

%\section{Low temperature transport} 

%\section{AFM/SEm Sample characterization} 

%Hauteur ??? 

\begin{table}
        \begin{center}
                \begin{tabular}{|c|c|}
\hline
		  \# of substrates & 12  \\
\hline
		  \# of FIB  slits & $\simeq$ 100 \\
\hline
                  \# of substrates with visible $\lambda$ DNA   & 5  \\
\hline
		  \# of substrates with conducting slits after $\lambda$ deposition  & 3  \\
		  
		  \hline
		  \# of slits on these  3 substrates  & 15    \\
\hline
		  \# of conducting slits after $\lambda$ deposition & 11  \\
\hline
                  
                  \# of slits on the control sample & 14 \\
\hline
		  \# of conducting slits after buffer & 0 \\
\hline   
		\end{tabular}
        \caption{Success rates for the formation of conductive junctions by deposition of $\lambda$ molecules. }
        \label{tab:Samples}
        \end{center}
\end{table}


\begin{thebibliography}{99}
\bibitem{dna} The $\lambda$ DNA solution   $5\;{\rm \mu g/ ml}$ was prepared with a commercial solution: "Invitrogen Cat no. 25250-028"  $250\;{\rm \mu g/ ml}$ in Tris-HCl     ${\rm (HOCH_2)_3 CNH_2 + H Cl}$ (pH 7.4), 10 mM  ${\rm Na Cl}$ and 5 mM EDTA    0.1 mM diluted in a magnesium chloride buffer ${\rm C H_3 C OO^{-} + N H_4^{+}} $  15 mM and  ${\rm Mg Cl_2}  $  5 mM.
\bibitem{orsaymoscou}Plasma functionalization and DNA deposition were attempted at two different laboratories, 
but conducting samples were obtained only in the setup in Moscow. Thus electrical measurements were performed in Orsay several days later.
\end{thebibliography}
\end{document}